\documentclass[11pt]{article}
\usepackage{jcappub}
\usepackage{bm}
\usepackage{color}
\usepackage{array}
\usepackage{graphicx}
\usepackage{multirow}
\usepackage{booktabs} 
\newcolumntype{P}[1]{>{\centering\arraybackslash}p{#1}}
\newcolumntype{M}[1]{>{\centering\arraybackslash}m{#1}}

\newcommand{\be}{\begin{equation}}
\newcommand{\ee}{\end{equation}}

\newcommand{\een}{\end{subequations}}
\newcommand{\ben}{\begin{subequations}}

\newcommand{\lsim}{\mathrel{\mathop{\kern 0pt \rlap
      {\raise.2ex\hbox{$<$}}}\lower.9ex\hbox{\kern-.190em $ \sim$}}}
\newcommand{\gsim}{\mathrel{\mathop{\kern 0pt
      \rlap{\raise.2ex\hbox{$>$}}}\lower.9ex\hbox{\kern-.190em $\sim$}}}

\title{Probing Sub-GeV Dark Matter with the Migdal Effect at JUNO}

\author[~]{Stefano Scopel,}
\author[~]{Gaurav Tomar}
\emailAdd{scopel@sogang.ac.kr}
\emailAdd{tomar@sogang.ac.kr}
\affiliation[~]{Department of Physics, Sogang University, 35 Baekbeom-ro, Mapo-gu, Seoul, 121-742, South Korea}
\affiliation[~]{Center for Quantum Spacetime, Sogang University, 35 Baekbeom-ro, Mapo-gu, Seoul, 121-742, South Korea}

\abstract{
We discuss the sensitivity of the JUNO neutrino detector to the Migdal ionization signal triggered by nuclear scattering events produced by sub--GeV weakly interacting massive particles (WIMPs). Exploiting JUNO's large target mass and the annual modulation effect we find that the aggregate rate from many independent and indistinguishable WIMP events in JUNO's liquid scintillator can be isolated  from the total dark rate of the  photomultipliers, potentially providing for spin-dependent interactions a world-leading sensitivity across the sub-GeV mass range.}

\begin{document}
\hspace*{107.5mm}{CQUeST-2026-0788}\\
\maketitle
\section{Introduction}
\label{sec:introduction}

Historically, Dark Matter (DM) searches have predominantly focused on weakly interacting massive particles (WIMPs) in the GeV-TeV mass  range~\cite{Jungman:1995df,Bergstrom:2000pn,Bertone:2004pz,Feng:2010gw}. 
After extensive experimental efforts the absence of any detection has significantly strengthened the motivation to explore lighter WIMPs, particularly in the sub-GeV mass range, leading to a major shift in both theoretical and experimental directions~\cite{Vergados:2005dpd,Moustakidis:2005gx,Bernabei:2007jz,Essig:2011nj,Battaglieri:2017aum,Ibe:2017yqa,Dolan_2017,Bell_2019,Essig_2019,Baxter:2019pnz,Liang:2019nnx,Flambaum_2020,Liu:2020pat,Knapen_2020,Liang:2020ryg,Wang_2021,Li:2022acp,Cox:2022ekg,Liang_2022,Blanco:2022pkt,Tomar:2022ofh,Adams:2022zvg,Berghaus:2022,Gu:2023pfg,Xu:2023wev,Kang:2024kec,Nakano:2024oon,Esposito:2025,Berghaus:2026kmj,NellenMondragon:2026thr,SENSEI:2023zdf,DAMIC-M:2025luv,PandaX:2025rrz,SENSEI:2025qvp,DAMIC-M:2025ltz,XENON:2026qow}.

In conventional direct-detection (DD) experiments, DM-nucleus scattering produces nuclear recoils (NR) with typical energies in the keV range, requiring extremely sensitive detectors. For sub-GeV DM, the expected recoil energies fall below this threshold, rendering standard nuclear recoil searches ineffective. To circumvent this limitation, several strategies have been proposed to enhance the observable energy deposition. These include boosted DM scenarios~\cite{Agashe:2014yua, Bringmann:2018cvk}, which are particularly well suited for searches in neutrino observatories with MeV-scale energy thresholds~\cite{Berger:2014sqa,Kopp:2015bfa,Bramante:2018tos,Ema:2018bih,Cappiello:2019qsw,PROSPECT:2021awi,Aoki:2023tlb,Dutta:2024kuj,Diurba:2025lky,Bhalla:2025vnq}, as well as DM-electron scattering (for comprehensive reviews, see e.g.~\cite{Knapen:2017xzo, Kahn:2021ttr, Essig:2022dfa, Hochberg:2022apz}). Another powerful approach is the use of the Migdal effect~\cite{migdal_1941}, where nuclear recoils are accompanied by the ionization of bound electrons, effectively converting sub-keV nuclear recoils into detectable electronic signals~\cite{XENON_migdal,ds50_migdal,cosine_migdal,supercdms_migdal,CDEX_2019,LUX_2018,EDELWEISS_2019,EDELWEISS_2022}.

Recently, Ref.~\cite{Leane:2025efj} proposed a novel approach to search for Sub-GeV DM using large-volume neutrino detectors, focusing specifically on DM-electron scattering and electronic excitations in scintillator molecules, with a detailed analysis performed for the JUNO detector. Subsequently, this framework was extended to other liquid scintillator targets and to include both electronic excitations and ionization~\cite{Santos-Olmsted:2025nuk}. In these scenarios, DM interactions with electrons can produce very low-energy electronic signals that lead to single-photomultiplier tubes (PMTs) hits. Although individual events are indistinguishable from the PMTs dark noise, the aggregate rate can be statistically extracted by exploiting the expected annual modulation of the DM signal. This method exploits the enormous target mass of neutrino detectors. The presence of eV-scale electronic excitation and ionization thresholds in scintillators enables otherwise inaccessible sensitivity to sub-GeV DM.

Motivated by the recent proposal of~\cite{Santos-Olmsted:2025nuk,Leane:2025efj}  
in the present paper we wish to investigate whether such novel approach can also probe sub-GeV DM through nuclear recoils via the Migdal effect. Specifically, in this work, we study the Migdal ionization signal induced by WIMP--nucleus scattering events in the JUNO detector and evaluate its potential sensitivity to sub-GeV DM masses. We compute the Migdal ionization rates for both spin-independent (SI) and spin-dependent (SD) DM-nucleon interactions using the publicly available package WimPyDD~\cite{wimpydd,wimpyc}. 
In our calculation we will both use the isolated--atom ionization probabilities for carbon and hydrogen available in the literature~\cite{Ibe:2017yqa} and approximate the molecular ionization probabilities using appropriately distorted Slater–type radial wave functions (see Appendix~\ref{app:lab_molecule}). The two approaches yield similar results. 

The work is organized as follows. In Sec.~\ref{sec:juno}, we review the JUNO experiment. The Migdal signal rate is described in Sec.~\ref{sec:sr}. Our analysis is presented in Sec.~\ref{sec:analysis}, and our conclusions are given in Sec.~\ref{sec:conclusion}. In the Appendices we provide a detailed discussion of our calculation of the ionization probability in hydrogen (\ref{app:iprob_hydrogen}) and carbon (\ref{app:iprob_carbon}), in particular addressing the issue of adapting existing isolated--atom approches to the case of large molecules. A few comments about the quenching factors adopted in our analysis are given in Appendix~\ref{app:quenching}.
\section{JUNO}
\label{sec:juno}
The Jiangmen Underground Neutrino Observatory (JUNO) is a liquid scintillator detector situated at a depth of 700 m underground that has recently started data taking~\cite{JUNO:2025fpc,JUNO:2025gmd}. Its 20 kton target mass and excellent energy resolution make it highly suitable for neutrino and astroparticle physics~\cite{JUNO:2020xtj,JUNO:2021vlw}. The detector is housed in a spherical acrylic vessel with a diameter of about 35 m surrounded by 17,612 20-inch PMTs and 25,600 3-inch PMTs, achieving an optical coverage of about 78$\%$~\cite{JUNO:2021vlw}. The primary goal of JUNO is the determination of the neutrino mass ordering through precise measurements of reactor antineutrino oscillations from multiple nuclear power plants at a baseline of about 53 km.

JUNO is primarily composed of linear alkylbenzene (LAB), containing carbon and hydrogen nuclei. Its chemical formula is ${\rm C}_6 {\rm H}_5 {\rm C}_n {\rm H}_{2n+1}$ (see Fig.~\ref{fig:lab_molecule} and Appendix~\ref{app:lab_molecule}) with $n=10-13$, doped with fluor (PPO) and wavelength shifter (bis-MSB)~\cite{Lombardi:2019epz}. To be conservative we take $n=10$ when determining the number of atoms in each LAB molecule. The total raw dark count rate in in JUNO is approximately $6\times 10^8$ Hz~\cite{JUNO:2022hlz,Zhang:2023dha}.
\section{Migdal Signal Rate}
\label{sec:sr}

Most existing studies of the Migdal effect are done in the standard free-atom approximation. On the other hand, recent studies of the Migdal effect in molecules are limited to a few simple toy models of diatomic molecules~\cite{Blanco:2022pkt}. In particular, a dedicated calculation of the molecular Migdal effect for complex organic molecules such as LAB is not currently available.
Indeed, in the specific examples of Ref~\cite{Blanco:2022pkt} it is shown that nuclear scattering in molecules can induce additional contributions beyond the isolated-atom picture, including non-adiabatic couplings and anisotropic responses associated with chemical bondings. Such effects arise from the coupling to internal molecular degrees of freedom, including vibrational and rotational modes, and are absent in purely atomic treatments. They may therefore modify both the spectral shape and the overall rate of low-energy ionization.

In absence of a full treatment, in our work we will compute the ionization probabilities for the Migdal effect adopting the standard simplified approach of Slater-type orbitals~\cite{Slater:1930,Zener:1930} with screening constants appropriately tuned to capture the distortion of the electronic wave functions in molecular bonds, and fixing ionization energies to observed values (see Appendix~\ref{app:lab_molecule} for details). In order to do so, we will confine our analysis to a range of momentum transfers where the WIMP--nucleus interaction is localized at the atomic level. The localization of the interaction is controlled by the typical momentum transfer $q \sim m_\chi v$, with $v \sim 10^{-3}$. When the associated de Broglie wavelength $\lambda \sim 1/q$ is smaller than the characteristic interatomic separation in the molecule, the scattering process becomes spatially localized at individual atomic centers, and interference effects among different atoms are suppressed. In this regime, the total ionization probability can be approximated as an incoherent sum over atomic contributions. 
For LAB, taking a representative interatomic distance $a \sim 1\text{--}4~\text{\AA}$, one has $1/a \sim 0.5\text{--}2~\text{keV}$. Then, requiring $q \gtrsim 1/a$ implies,
\begin{equation}
q \sim m_\chi v \gtrsim (0.5\text{--}2)~\text{keV} \,,
\end{equation}
which translates into a lower bound on the DM mass,
\begin{equation}
m_\chi \gtrsim (0.5\text{--}2)~\text{MeV} \,.
\end{equation} 

\noindent Such momentum range also ensures that the DM-nucleus collisions happen rapidly compared to the time scale set by the potential well of the molecular vibrations (impulse approximation~\cite{Knapen_2020, Ibe:2017yqa, Essig_2019, Cox:2022ekg}). In the following we will discuss our results in the safe range $m_\chi\gsim$ 10 MeV.  

When a WIMP scatters off a nuclear target inside JUNO it can excite or ionize it, producing photons that are detected by the PMTs. The total event rate is then given by,
\be
 R_{\rm tot}=R_{\rm exc}+R_{\rm ion},
\ee
where $R_{\rm exc}$ denotes the event rate due to electronic excitation, while $R_{\rm ion}$ is the ionization event rate. Since the probability for electron excitation in the Migdal process is significantly smaller than that for ionization, the contribution of the former to the total event rate can be safely neglected~\cite{Ibe:2017yqa}, leading to,
\be
 R_{\rm tot}=R_{\rm ion}.
\ee
The differential ionization event rate in a direct detection experiment arising from the Migdal effect is given by~\cite{Ibe:2017yqa},
\begin{equation}
\label{eq:diff_rate_migdal0}
\frac{dR_{\rm tot}(t)}{dE_{\rm det}}
=
\epsilon(E_{\rm det})\,\sum_T
\int_0^{\infty} dE_R
\int_{v_{\min}(E_R)}^{\infty} dv_T\;
\frac{d^3 R_{\chi T}(t)}{dE_R\, dv_T\, dE_{\rm det}} \, .
\end{equation}
The triple-differential rate factorizes into a nuclear-recoil contribution and an electronic ionization probability, 
\begin{equation}
\label{eq:diff_rate_migdal}
\frac{d^3 R_{\chi T}(t)}{dE_R\, dv_T\, dE_{\rm det}}
=
\frac{d^2 R_{\chi T}(t)}{dE_R\, dv_T}
\times
\sum_{s}
\frac{d}{dE_e}
p^c_{q_e}\!\left(s \rightarrow E_e\right) \, ,
\end{equation}

\noindent where the velocity and recoil-energy differential scattering rate is given by,
\begin{equation}
\label{eq:diff_rate_NR}
\frac{d^2 R_{\chi T}(t)}{dE_R\, dv_T}
=
N_T\,
\frac{\rho_\chi}{m_\chi}\,
v_T\,
f(v_T,t)\,
\frac{d\sigma_T}{dE_R} \, .
\end{equation}
In the equation above $m_\chi$ denotes the WIMP mass, $\rho_\chi$ is the local DM mass density in the Solar neighborhood, and $N_T$ is the number of target nuclei of species $T$ in the detector. The quantity $d\sigma_T/dE_R$ denotes the differential WIMP–nucleus scattering cross section, whose explicit form is specified below in the non-relativistic effective field theory (NREFT) framework. Here, $v_T$ = $|\mathbf{v}_T|$ is the WIMP speed in the lab frame (i.e. relative to the target nucleus assumed at rest), and $f(v_T,t)$ denotes the corresponding velocity distribution, which is obtained by averaging the boosted Galactic-frame velocity distribution $f_{\rm gal}$ over incoming WIMP angles,
\begin{equation}
\label{eq:speed_dist}
f(v_T,t)
=
v_T^2
\int d\Omega\,
f_{\rm gal}\!\left(\mathbf u=\mathbf v_T+\mathbf v_E(t)\right).
\end{equation}
\noindent In the equation above $\mathbf u$  and $\mathbf{v}_E(t)$ are the velocities of the WIMP and of the Earth in the Galactic rest frame and $\Omega$ is the angle between $\mathbf{v}_T$ and $\mathbf{v}_E(t)$. For $f_{\rm gal}$ we adopt a standard Maxwell--Boltzmann distribution truncated at the escape speed, $|\mathbf{u}|\le u_{\rm esc}$,

\begin{eqnarray} 
f_{\rm gal}(\mathbf u) &=& \frac{1}{N}\left(\frac{3}{2\pi v^2_{\rm rms}}\right)^{3/2} e^{-\frac{3|\mathbf u|^2}{2v^2_{\rm rms}}}\, \Theta(u_{\rm esc}-|\mathbf u|),\nonumber\\ N &=& \left({{\rm erf} \left(\frac{3u_{\rm esc}^2}{2v_{\rm rms}^2} \right)}-\frac{2}{\sqrt{\pi}}\frac{3u_{\rm esc}^2}{2v_{\rm rms}^2}e^{-\left(\frac{3u_{\rm esc}^2}{2v_{\rm rms}^2}\right)^2}\right), \end{eqnarray}

\noindent with $v_{\rm rms}=\sqrt{3/2}v_0$ and $v_0$ the Galactic rotational velocity. In our analysis we take $v_0=220$ km/s~\cite{green} and $u_{\rm esc}=550~\mathrm{km/s}$~\cite{vesc, vesc_2014}.

The time dependence of the event rate (Eq.~(\ref{eq:diff_rate_migdal0})) arises from the Earth's velocity with respect to the Galactic frame~\cite{Freese_2012},

\begin{equation}
 \label{eq:vearth_vec}
 \mathbf{v}_E(t) = \mathbf{v}_\odot + v_{\text{orb}} \left[ \hat{\mathbf{e}}_1 \cos \left( \frac{2\pi}{1\text{ yr}} (t - t_0) \right) + \hat{\mathbf{e}}_2 \sin \left( \frac{2\pi}{1\text{ yr}} (t - t_0) \right) \right].
\end{equation}

\noindent In the equation above $t_0=$June 2$^{nd}$ denotes the time at which the speed of Earth relative to the Galactic frame is maximal, $v_{\rm orb}\simeq29~\mathrm{km/s}$, and, in Galactic coordinates,  $\mathbf v_{\odot}\simeq (11.1, v_0+12~\mathrm{km/s},7.3)$ km/s (taking into account the peculiar component of the solar system with respect to the Galactic rotation) while $\hat{\mathbf{e}}_1$ and $\hat{\mathbf{e}}_2$ are two orthonormal vectors lying in the ecliptic plane with $\hat{\mathbf{e}}_1$ pointing in the direction of Earth's velocity at time $t_0$. In particular, the relevant quantity for phenomenology is the norm of $\mathbf{v}_E$,   

\begin{equation}
\label{eq:vearth}
v_E(t)=|\mathbf v_E(t)|
=
v_{\odot}
+
v_{\rm orb}\cos\gamma\,
\cos\!\left(\frac{2\pi}{1~{\rm yr}}(t-t_0)\right),
\end{equation}
\noindent with $v_{\odot}$ = $|\mathbf v_{\odot}|$ and $\cos\gamma\simeq0.49$ accounts for the inclination of the ecliptic plane relative to the Galactic plane. 

Finally, in Eq.~(\ref{eq:diff_rate_migdal}) $p^c_{q_e}(s\rightarrow E_e)$ denotes the ionization probability for an electron initially occupying the orbital $s$ of the LAB molecule, evaluated at $q_e=m_e\sqrt{2E_R/m_T}$, where $E_R$ is the nuclear recoil energy, $m_T$ is the nuclear mass, and $m_e$ is the electron mass. In the following, we use both the isolated--atom ionization probabilities available in the literature~\cite{Ibe:2017yqa} and approximate the molecular ionization probabilities using appropriately distorted Slater–type radial wave functions (see Appendix~\ref{app:lab_molecule}). 

It is worth mentioning that, when boosted in the laboratory frame, the ionization probabilities $p^c_{q_e}$ are enhanced by the recoil of the nucleus and increase linearly with $E_R$. 

In Eq.~(\ref{eq:diff_rate_migdal0}) $E_{\rm det}$ represents the total deposited energy given by,

\begin{eqnarray}\label{eq:edet}
E_{\rm det}&=&QE_R+E_{\rm EM},\nonumber\\
    E_{\rm EM}&=&E_e+E_{ion},
\end{eqnarray}
\noindent where $Q$ is the quenching factor (i.e. the fraction of the nuclear recoil energy converted into scintillation), $E_e$ denotes the energy of the outgoing electron and $E_{ion}$ is the atomic de-excitation energy released following ionization. In particular, in the following we will adopt $Q$ = 0.1 for WIMP scattering events off hydrogen and $Q\rightarrow$  0 for WIMP scattering events off carbon (see Appendix~\ref{app:quenching}). 
As already pointed out, when the WIMP mass is low enough the electromagnetic energy deposited by the Migdal ionization process eventually dominates over the quenched nuclear recoil contribution, i.e.  $Q E_R \ll E_{\rm EM}$ in Eq.~(\ref{eq:edet}).

Moreover, the minimum WIMP velocity required for the recoil of the nucleus at a given energy $E_R$ is given by,
\be
    v_{min}(E_R)=\frac{m_{T} E_R + \mu_{\chi T} E_{\rm EM}}{\mu_{\chi T}\sqrt{2m_T E_R}},
    \label{eq:vmin}
\ee
where $\mu_{\chi T}$ represents the WIMP--nucleus reduced mass. 

Due to the small momentum transfer, the WIMP-nucleus scattering process is well described within the NREFT framework~\cite{haxton1,haxton2,all_spins}. The most general Galilean-invariant interaction Hamiltonian can be written as,
\begin{equation}
\mathcal{H}(\mathbf{r})=\sum_{\tau=0,1}\sum_{j} c_j^\tau\,\mathcal{O}_j(\mathbf{r})\,t^\tau ,
\label{eq:H}
\end{equation}
where $t^0=1$ and $t^1=\tau_3$ are the isoscalar and isovector operators.
In general, at linear order in the incoming WIMP particle speed $v$ spin-$1/2$ WIMP--nucleus scattering admits 14 independent operators ~\cite{haxton1,haxton2,all_spins}. In this work we restrict ourselves to the standard SI and SD interactions, that are the only non-vanishing ones at zero order in $v$, and that in the standard notation of~\cite{haxton1} correspond to the operators $\mathcal{O}_1$ and $\mathcal{O}_4$, respectively. 
In this case the effective Hamiltonian of Eq.~(\ref{eq:H}) reduces to,
\begin{equation}
\mathcal{H}(\mathbf{r}) =
\sum_{\tau=0,1}\left[
c_1^\tau\,\mathcal{O}_1(\mathbf{r})
+ c_4^\tau\,\mathcal{O}_4(\mathbf{r})
\right] t^\tau,
\end{equation}
where $c_1^\tau$ and $c_4^\tau$ denote the strengths of the non-relativistic operators
$\mathcal O_1 = \mathbf{1}_\chi\,\mathbf{1}_N$
and
$\mathcal O_4 = \vec{S}_\chi \cdot \vec{S}_N$, respectively. 

The DM-nucleus differential cross section entering Eq.~(\ref{eq:diff_rate_NR}) is given by,
\begin{equation}
\frac{d\sigma_T}{dE_R}
=
\frac{2 m_T}{4\pi v_T^2}
\left[
\frac{1}{(2j_\chi + 1)(2j_T + 1)}
\left| \mathcal{M}_T \right|^2
\right],
\end{equation}
where the squared matrix element averaged over initial DM and nuclear spins $j_\chi$ and $j_T$, respectively, can be expressed as,

\begin{equation}
  \frac{1}{2j_{\chi}+1} \frac{1}{2j_{T}+1}\sum_{spin}|{\cal M}_T|^2=
  \frac{4\pi}{2j_T+1}\sum_{\tau\tau^{\prime}}\sum_l R_l^{\tau\tau^{\prime}} (q,v)W_{l,T}^{\tau\tau^{\prime}}(q).
\label{eq:haxton_40}
\end{equation}

\noindent In the expression above the WIMP and nuclear physics are
factorized in the two $R_l^{\tau\tau^{\prime}}$ and
$W_{l,T}^{\tau\tau^{\prime}}$ functions, respectively, with
$\tau,\tau^{\prime}=0,1$ the nuclear isospin. Explicit expressions of the WIMP response functions
$R_l^{\tau\tau^{\prime}}$ are provided in~\cite{haxton2}
and~\cite{all_spins}. On the other hand the $W_{l,T}^{\tau\tau^{\prime}}$ are nuclear response functions where
$l$=$M$,$\Sigma^{\prime\prime}$
,$\Sigma^{\prime}$,$\Phi^{\prime\prime}$, $\Phi^{\prime\prime}M$,
$\tilde{\Phi}^{\prime}$,$\Delta$, $\Delta\Sigma^{\prime}$ represent
one of the possible nuclear interaction types, in the single--particle
interaction limit~\cite{haxton1,haxton2}\footnote{In the case of a SI interaction the nuclear recoil is driven by the $W_{M,T}$, while the SD process depends of $W_{\Sigma^{\prime},T}$ and $W_{\Sigma^{\prime\prime},T}$.}. For the operators considered here, the nuclear response functions are taken from Refs.~\cite{haxton1,haxton2}.

It is convenient to express the couplings $c_1^N$ and $c_4^N$ ($N=p,n$) in terms of the corresponding WIMP-nucleon cross sections,
\begin{align}
\sigma_{\rm SI}^N &= \frac{(c_1^N)^2\,\mu_{\chi N}^2}{\pi}, \\
\sigma_{\rm SD}^N &= \frac{3}{16}\frac{(c_4^N)^2\,\mu_{\chi N}^2}{\pi},
\end{align}
where $\mu_{\chi N}$ is the DM--nucleon reduced mass. The proton and neutron couplings are related to the isoscalar and isovector coefficients via
$c_i^p=(c_i^0+c_i^1)/2$ and $c_i^n=(c_i^0-c_i^1)/2$, with $i=1,4$. In the following, for both the SI and the SD case we will assume an isoscalar interaction, i.e. $c_i^1$ = 0  and $c_i^p$ = $c_i^n$, and set $\sigma_{\rm SI,SD}$ = $\sigma_{\rm SI,SD}^0 \equiv 4\sigma_{\rm SI,SD}^p$.

The corresponding Migdal differential rate $\frac{d^2R_{\chi T}}{dE_R dv_T}$ in Eq.~(\ref{eq:diff_rate_migdal}) is computed following Ref.~\cite{sogang_scaling_law_nr}, as implemented in the publicly available package WimPyDD~\cite{wimpydd,wimpyc}. The rate scales linearly with the local DM density, for which we take $\rho_\chi=0.3~\mathrm{GeV/cm^3}$.
\begin{figure}
   \centering
       \centering
       \includegraphics[width=0.49\linewidth]{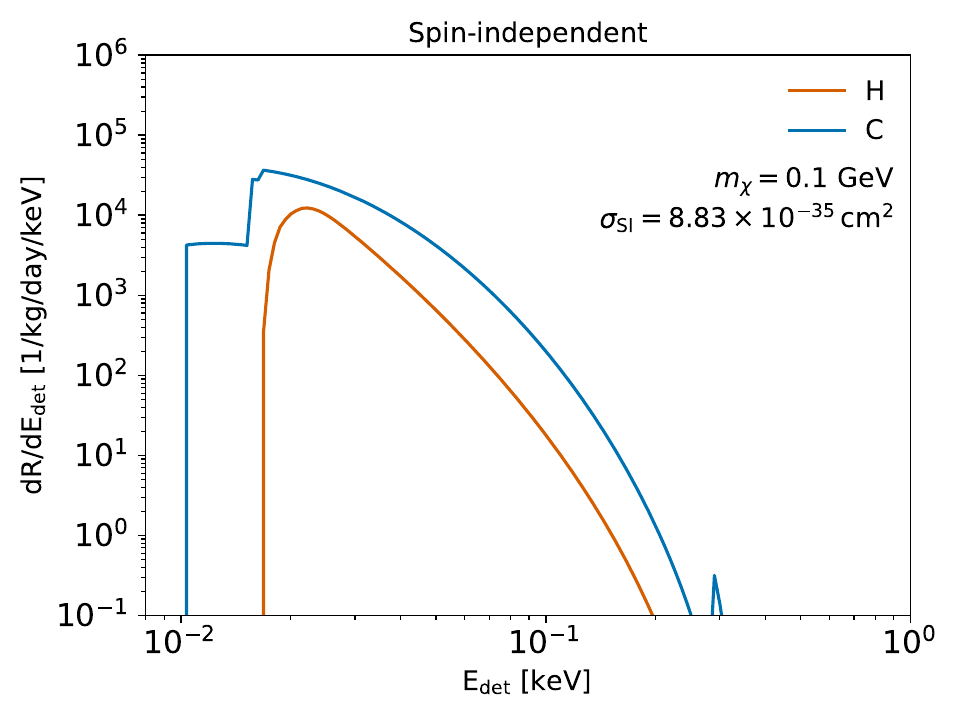}
       \includegraphics[width=0.49\linewidth]{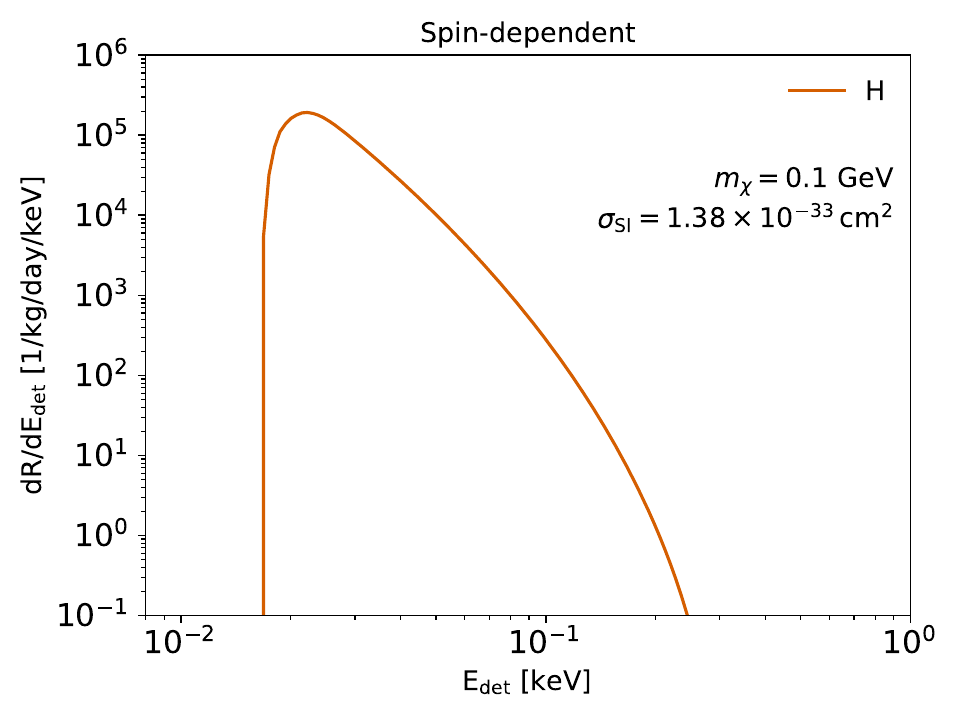}
   \caption{The Migdal differential rate as a function of detected energy is shown for hydrogen and carbon targets in LAB, for $m_\chi = 0.1~\mathrm{GeV}$. The left and right panels correspond to SI and SD interactions, respectively.
}
   \label{fig:diff_rate_migdal_comb}
\end{figure}
\begin{figure}
    \centering
        \centering
        \includegraphics[width=0.6\linewidth]{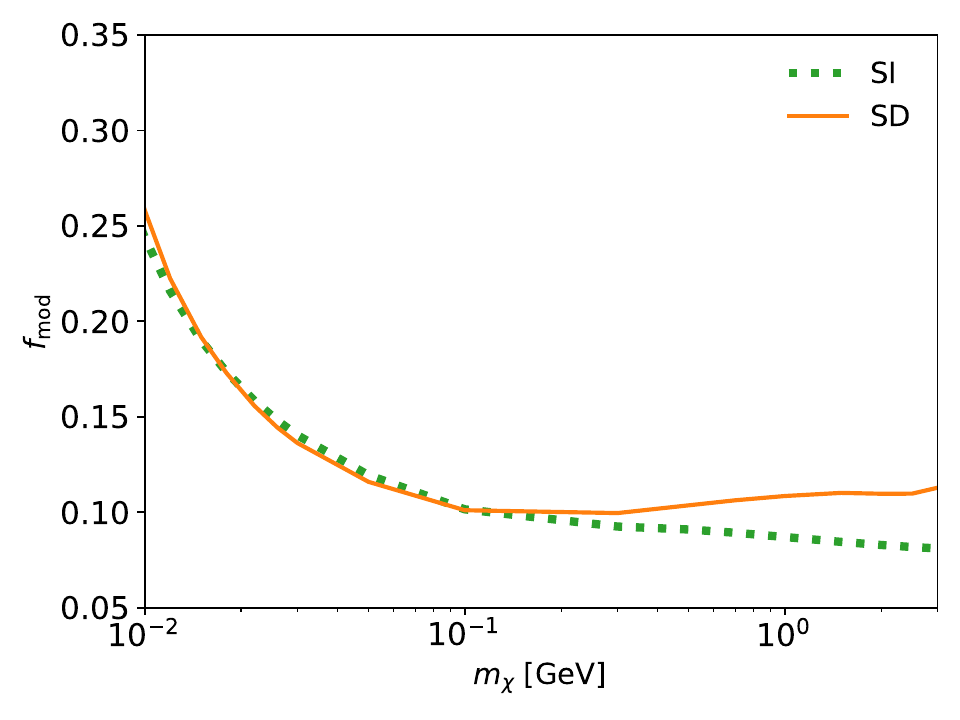}
    \caption{The modulation fraction $f_{\rm mod}$ (Eq.~(\ref{eq:fmod})) as a function of $m_\chi$ for SI and SD interactions is presented with dashed and solid lines respectively.}
    \label{fig:fmod}
\end{figure}
\section{Analysis}
\label{sec:analysis}
JUNO is sensitive to scintillation signals produced by excitation and ionization of LAB molecules. Scintillation is also produced directly from the nuclear recoil, by the same mechanism used by other DD searches such as XENONnT~\cite{XENON:2026qow} or LZ~\cite{LZ}. However, at low WIMP masses ($m_\chi\lesssim$ 1 GeV) the recoil energy is low  and this contribution is suppressed (see later and Appendix~\ref{app:quenching}). Taking advantage of the huge detector volume (20~kton) and low electronic excitation thresholds of order $\mathcal{O}(10)\,\mathrm{eV}$, Ref.~\cite{Leane:2025efj} pointed out that the dark count rate of the PMTs at JUNO can be used to search for a DM signal. 

The enormous dark count rate at JUNO, $\mathcal{O}(10^{16})$ events per year, severely limits the sensitivity to a reliable DM signal. However, the motion of the Earth around the Sun induces an annual modulation in the DM flux, which can be used to distinguish a potential DM signal from most background sources~\cite{Leane:2025efj}. The annual modulation of the event rate is parametrized as, 
\be
R(t)=\frac{R_{\rm tot}}{T}\left[1+f_{\rm mod}\cos\!\left(\frac{2\pi}{1~{\rm yr}}(t-t_0)\right)\right],
\label{eq:rate_td}
\ee
where $R_{\rm tot}$ denotes the total number of Migdal events expected over the observation period $T$, 
so that $R_{\rm tot}/T$ corresponds to the time-averaged event rate. Here $t=0$ corresponds to 2 March, when the Earth's orbital velocity has no component along the Sun's motion in the Galactic frame, while the detection rate peaks around 2 June and reaches a minimum around 2 December. In the equation above, $f_{\rm mod}$ is the modulation amplitude defined as,
\be
f_{\rm mod}=\frac{R({\scriptstyle \rm Jun})-R({\scriptstyle \rm Dec})}{R({\scriptstyle \rm Jun})+R({\scriptstyle \rm Dec})},
\label{eq:fmod}
\ee
where $R({\scriptstyle \rm Jun})$ and $R({\scriptstyle \rm Dec})$ are evaluated using Earth velocities $v_E=247~\mathrm{km/s}$ and $v_E=217~\mathrm{km/s}$, corresponding to 2 June and 2 December, respectively. 
 
As discussed in Refs.~\cite{Leane:2025efj,Santos-Olmsted:2025nuk}, the DM signal can be distinguished by fitting the binned time-series data for the observed event counts
$N(t)=B(t)+R(t)$, where $B(t)=B_{\rm tot}/T$ denotes the time-averaged
background rate, with $B_{\rm tot}$ the total background accumulated over
the exposure time $T$, and $R(t)$ is the DM signal rate given in Eq.~(\ref{eq:rate_td})\footnote{With the goal to assess JUNO's potential sensitivity we assume a constant background. In a real experiment the raw measured dark rate may not be constant. See Ref.~\cite{Leane:2025efj} for a discussion of the related challenges of real data.}. The amplitude of the modulated signal component is $R_{\rm tot} f_{\rm mod}$, while the statistical background fluctuation is given by $\sqrt{B_{\rm tot}}$. Requiring a $95\%$ C.L. sensitivity to the
modulated signal yields,
\be\label{eq:bound}
R_{\rm tot} = \frac{2\sqrt{B_{\rm tot}}}{f_{\rm mod}}.
\ee
For a background of $B_{\rm tot}\simeq2\times10^{16}$ events per year at JUNO,
we obtain a conservative estimate of
$R_{\rm tot}\sim3\times10^{9}$ events per year.

We computed the Migdal ionization rate of Eq.~(\ref{eq:diff_rate_migdal}) using WimPyDD~\cite{wimpydd,wimpyc}, incorporating the ionization probabilities calculated in Appendix~\ref{app:lab_molecule}. As mentioned before, our calculation considers both SI and SD isoscalar interactions, assuming a standard Maxwellian velocity distribution for DM. The resulting differential rates in the LAB molecule are shown in Fig.~\ref{fig:diff_rate_migdal_comb}, where only hydrogen contributes to SD interactions, since carbon does not have nuclear spin. 

We estimate the potential sensitivity of JUNO by calculating the projected 95\% C.L. exclusion plot on $\sigma_{\rm SI}$ and $\sigma_{\rm SD}$ for the non-observation of a Migdal--induced modulation signal integrated over its full energy range. For the latter we assume $E_{\rm det}=0.14\text{--}10$ keVee taking into account the dark count rate threshold (0.25 photo-electrons) from~\cite{juno_threshold}, and the observed photoelectron yield factor $\epsilon(E_{\rm det})=1.7 \times (E_{\rm det}/{\rm keVee})$ photo-electrons (see Table~I in~\cite{Santos-Olmsted:2025nuk}). 
The modulation amplitude $f_{\rm mod}$ integrated over such energy range is plotted in Fig.~\ref{fig:fmod} as a function of $m_\chi$ for the SI (dashed line) and the SD (solid-line) interactions. For the interactions considered, $f_{\rm mod}$ varies in the range $10$--$25\%$.

\begin{figure}
    \centering
        \centering
        \includegraphics[width=0.49\linewidth]{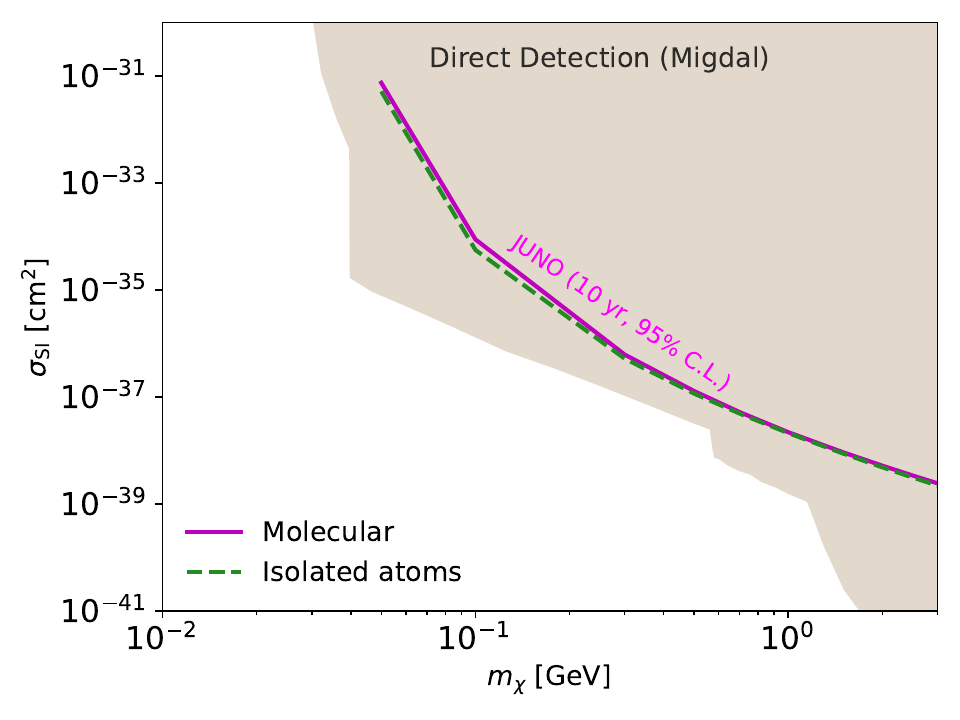}
        \includegraphics[width=0.49\linewidth]{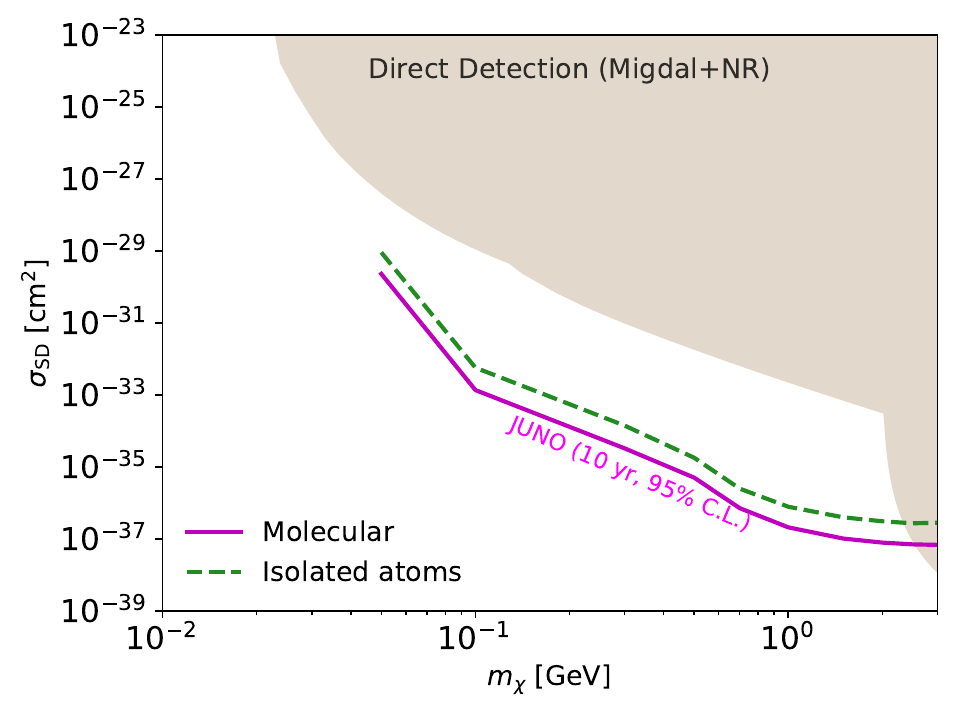}
    \caption{Projected $95\%$ C.L. sensitivity of JUNO to Migdal-induced scattering events shown as the projected exclusion plot to the isoscalar SI (left) and SD (right) WIMP--nucleon cross sections, $\sigma_{\rm SI}$ and $\sigma_{\rm SD}$, assuming no observation of a modulation signal and an exposure of 10 years. The shaded regions are excluded by present DD limits~\cite{XENON_migdal,supercdms_migdal,cosine_migdal, ds50_migdal, pico60_2019}. Solid lines are calculated with the molecular ionization probabilities obtained in Appendix~\ref{app:lab_molecule}, while dotted ones make use of the isolated--atom evaluations of Ref.~\cite{Ibe:2017yqa}.
}
\label{fig:cs_migdal}
\end{figure}

The results of our analysis are
 shown in Fig.~\ref{fig:cs_migdal}. In both plots shaded regions are excluded by present DD limits~\cite{XENON_migdal,supercdms_migdal,cosine_migdal, ds50_migdal, pico60_2019}. 
 Moreover, for the solid lines the Migdal effect is  calculated with the molecular ionization probabilities obtained in Appendix~\ref{app:lab_molecule}, while dotted ones make use of the isolated--atom ionization probabilities of Ref.~\cite{Ibe:2017yqa}. In particular, 
 for a SI interaction (driven by WIMP-carbon scattering events) we find that the sensitivity of JUNO is never competitive with existing bounds (in particular, with the bound from DarkSide-50~\cite{ds50_migdal}). On the other hand,  in the case of a SD interaction WIMP-hydrogen scattering events drive the JUNO sensitivity beyond that of existing DD experiments for 50 MeV$\lesssim m_\chi\lesssim$2 GeV (notice that in this case the DarkSide-50 bound is not present since argon has zero nuclear spin and has no sensitivity to SD interactions). Notice that our conclusions are the same irrespective on whether we calculate the Migdal effect making use of isolated--atom or molecular ionization probabilities. 

As already pointed out, in this work we discussed our results in the safe range $m_\chi\ge$ 10 MeV, where the LAB molecule can be treated as a collection of isolated carbon and hydrogen atoms. In order to extend the analysis to lower masses a more refined treatment of the Migdal molecular effect would be required. However,
such range is sufficient to demonstrate how JUNO can extend the sensitivity of existing Migdal experiments. 

\section{Conclusion}
\label{sec:conclusion}
In this work we have shown that the Migdal ionization radiation triggered by  WIMP-hydrogen scattering events off the huge number (20kton) of LAB molecular targets in the JUNO neutrino experiment can induce a visible annual modulation signal beyond the sensitivity of existing DD searches sensitive to a WIMP-nucleus SD interaction. On the other hand, the analogous signal produced by WIMP-carbon scattering events are not competitive to existing DD experiments searching for WIMP-nucleus SI interactions. Our conclusions assume in both cases an isoscalar interaction ($c_i^p$ = $c_i^n$) and are similar if we use isolated--atom or molecular ionization probabilities. 

A few comments are in order. In our calculation of the ionization probabilities we assumed impulse approximation and neglected interference effect among different atoms, modeling the distortion of atomic orbitals with Slater–type radial wave functions. While such approximations should be in general safe for the WIMP mass range that we considered, non–adiabatic effects might still be present at the lowest ionization energies (for instance, for the $p_z$ orbital in the aromatic ring). Moreover, while we simplistically assumed a time--independent dark noise in JUNO, in real experiments, the raw measured dark rates are
not constant~\cite{Borexino, SK}. In such case the dominant time--dependent background should be from muons, whose modulation phase is different compare to DM. In order to discriminate it a full time--projection analysis of the detected count rate should be required, besides the simple June-December estimation that we adopted. 

\section*{Acknowledgments}
This research was supported by Basic Science Research Program through the National Research Foundation of Korea (NRF) funded by the Ministry of Education through the Center for Quantum Spacetime (CQUeST) of Sogang University with grant number RS-2020-NR049598 and by the Ministry of Science and ICT with grant number RS-2025-24523022. GT acknowledges support from the Munich Institute for Astro-, Particle and BioPhysics (MIAPbP), which is funded by the Deutsche Forschungsgemeinschaft (DFG, German Research Foundation) under Germany’s Excellence Strategy – EXC-2094–390783311, and the support of the Technical University of Munich (TUM) during his visit, where part of this work was carried out. He also acknowledges the warm hospitality of the Reichard family during his stay in Munich.
\appendix
\section{Ionization probabilities in the LAB molecule}
\label{app:lab_molecule}

\begin{figure}
    \centering
        \centering
        \includegraphics[width=0.59\linewidth]{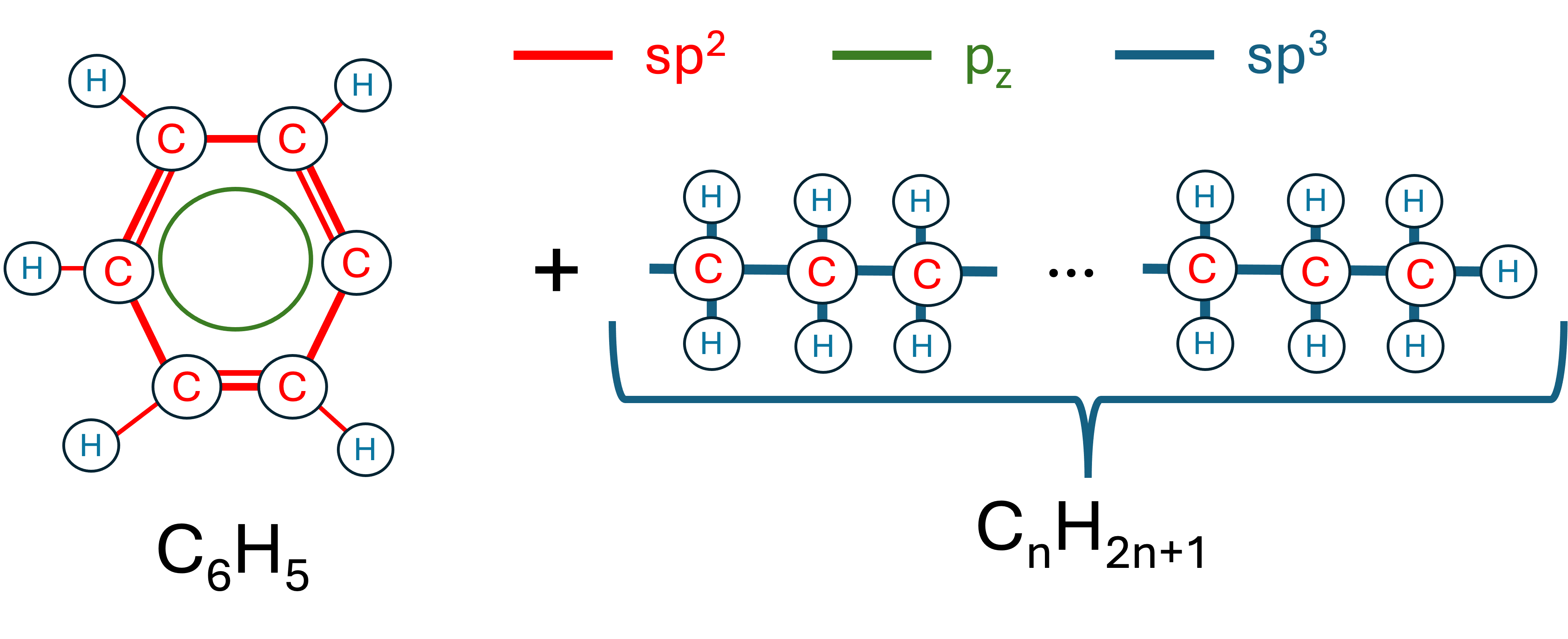}
\caption{
Schematic representation of the LAB molecule ($C_6H_5 C_{n} H_{2n+1}$). The different colors represent the types of hybridized orbitals: red: $sp^2$; blue: $sp^3$; green: $p_z$.}
    \label{fig:lab_molecule}
\end{figure}

A schematic view of the linear LAB molecule is provided in Fig.~\ref{fig:lab_molecule}. Its topology consists of a flat hexagonal benzene ring attached to a hydrocarbon chain. Each of the 6 carbon atoms in the ring is attached to two other carbons and a hydrogen through three hybridized $sp^2$ bonds with trigonal planar geometry and angles of 120$^{\circ}$. Moreover, each of the 6 atoms in the ring provides one additional electron to a de--localized $p_z$ orbital that extends above and below the plane of the ring (in Fig.~\ref{fig:lab_molecule} the $p_z$ orbital is shown with a green circle at the center of the ring)\footnote{The 6 $p_z$ electrons split further into 4 $p_z$(1,1') electrons and 2 $p_z(2)$ electrons (see Table~\ref{table:C_orbital} and Ref. \cite{Blanco:2019lrf}).}. On the other hand, from each carbon in the chain four $sp^3$ hybridized bonds are attached to two carbon and two hydrogen atoms with tetrahedral shape and angles of $\simeq 109.5^{\circ}$.

As already pointed out, attempts to calculate from first principles the Migdal effect in molecules are limited to the case of diatomic molecules~\cite{Blanco:2022pkt}. In particular, in the case of an isolated atom the ionization probability is driven by a dipole operator triggered by the mismatch between the reference frames of the nucleus and that of the electron.  However, it is worth pointing out that in Ref.~\cite{Blanco:2022pkt} it was shown that for the molecular excitation process at the lowest transition energies (the effect scales as the inverse of the fourth power of the latter~\cite{Blanco:2022pkt}) the dipole operator can get a sizeable non--adiabatic contribution, where the nuclear recoil can trigger an internal energy transition of the electron. Indeed, in Ref.~\cite{Esposito:2025} it was shown analytically that in a crystal (i.e., in a molecule of infinite size) the adiabatic contribution vanishes and only the non--adiabatic term survives, confirming the effective theory calculation in crystals of Ref.~\cite{Berghaus:2022}. The case of ionization in a molecule as large as the LAB one is intermediate between an isolated atom and a crystal, suggesting that, at least at the lowest ionization energies, the non--adiabatic effect might be sizeable. In our analysis we adopt a simplified standard approach, where the transition is assumed adiabatic, and the ionization probabilities are obtained using Slater--type radial wave functions~\cite{Slater:1930, Zener:1930},

\begin{equation}
    R_n=\frac{1}{\sqrt{(2n)!}} \left( \frac{2 Z_{eff}}{n a_0} \right)^{n + 1/2} r^{n-1} e^{-\frac{Z_{eff}r}{n a_0}},
    \label{eq:slater_orbital}
\end{equation}

\noindent where $n$ is the principal quantum number and $a_0$ is the Bohr radius. The screening constant, $Z_{eff}$, is tuned to reproduce the molecular distortion of the electron spatial distribution and the observed ionization energies, $E_{ion}$~\cite{hehre1969self,thompson2001xray,potts1972photoelectron,karlsson1973high}.
This implies that we may underestimate the ionization probability at the lowest ionization energies (i.e. the $p_z$ orbital).

In the case of a SD $\mathcal O_4 = \vec{S}_\chi \cdot \vec{S}_N$ WIMP--nucleus interaction the Migdal effect only probes the interaction of the DM particles with hydrogen targets, since carbon nuclei have no spin. On the other hand, if instead the WIMP--nucleus interaction is of the SI type the cross section is enhanced by the atomic mass number and the Migdal effect is driven by WIMP-carbon interactions. 

\subsection{Hydrogen ionization probability}
\label{app:iprob_hydrogen}
\begin{table}
\centering
\begin{tabular}{|c|c|c|}
\toprule
C-H bond & $Z_{eff} $ & $E_{ion}$ (eV) \\
\midrule
Aromatic ($sp^2$)& 1.38 & 18.5\\
Alkyl ($sp^3$)& 1.25 & 15.5\\
\bottomrule
\end{tabular}
\caption{Slater screening constants $Z_{eff}$ and ionization probabilities for molecular orbitals in hydrogen targets~\cite{potts1972photoelectron, karlsson1973high}}
\label{table:H_orbitals}
\end{table}

Carbon and hydrogen have similar electronegativities, with carbon slightly more electronegative than hydrogen. As a consequence, in the LAB molecule the C--H bonds are slightly polarized and almost covalent, shifting the electron density slightly toward carbon.
This is particularly relevant for the Migdal effect driven by WIMP-hydrogen scattering events, because it ensures that the C--H bond does not strip the hydrogen nucleus of its only electron. As a consequence, the hydrogen electron remains localized near the nucleus, although its binding energy and spatial distribution are perturbed by the presence of the carbon atom in the bond.  

The Migdal transition amplitude for hydrogen can then be written as,
\be
M^{E_e l' m'}_{nlm} = \int d^3\mathbf{r}\,
\psi^*_{E_e l' m'}(\mathbf{r}) \exp\!\left(i m_e\,\mathbf{v}\cdot\mathbf{r}\right) \psi_{nlm}(\mathbf{r}),
\ee
where $\mathbf{v}$ is the nuclear recoil velocity and $\mathbf{r}$ denotes the electron position operator relative to the nucleus. 
The quantum numbers $(n,l,m)$ specify the initial bound electron state, while $(l',m')$ and $E_e$ denote the angular momentum quantum numbers and energy of the final continuum electron state, respectively. 
Under the dipole approximation, valid when $m_e \mathbf{v}\!\cdot\!\mathbf{r} \ll 1$, the exponential can be expanded to first order in a Taylor series, yielding
\be\label{eq:amplitude}
M^{E_e l' m'}_{nlm} \simeq i m_e \int d^3\mathbf{r}\,
\psi^*_{E_e l' m'}(\mathbf{r}) \left(\mathbf{v}\cdot \mathbf{r}\right) \psi_{nlm}(\mathbf{r}).
\ee
\begin{figure}
    \centering
        \centering
        \includegraphics[width=0.6\linewidth]{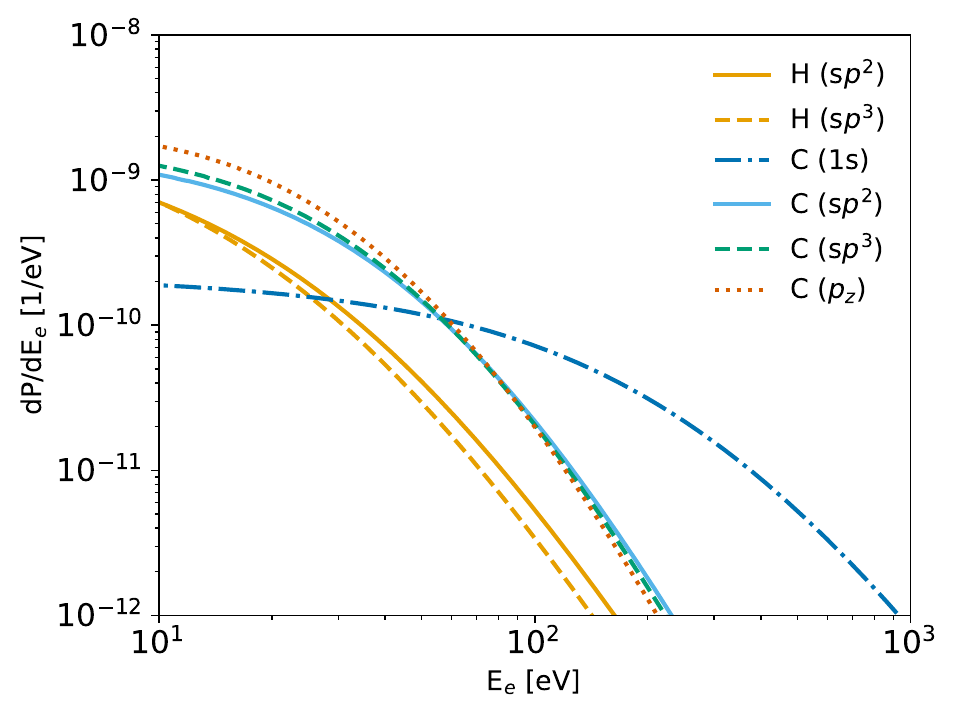}    
\caption{
Comparison of the Migdal ionization probability of hydrogen (Eq.~(\ref{eq:iprobh})) and the carbon ionization probabilities calculated in Ref.~\cite{Ibe:2017yqa} for $q_e = 1\,\mathrm{eV}$.}
    \label{fig:onization_prob}
\end{figure}
For the hydrogen ground state $(n=1,l=0,m=0)$ we simply adopt,
\[
\psi_{100}(\mathbf{r}) = R_{1}(r)\,Y_{00}(\theta,\phi)
= R_1(r)\,\frac{1}{\sqrt{4\pi}},
\]
\noindent i.e. we assume that the ionization probability is driven by the inner spherically symmetric part of the orbital closer to the nucleus and the radial wavefunction $R_1$ given in Eq.~(\ref{eq:slater_orbital}), with $Z_{eff}$ and the ionization energy provided in Table~\ref{table:H_orbitals} to take into account the distortion due to the molecular $sp^2$ and $sp^3$ bonds.
The continuum final state is,
\be
\psi_{E_e l' m'}(\mathbf{r}) = R_{E_e l'}(r)\, Y_{l' m'}(\theta,\phi),
\ee
with the Coulomb radial wave function~\cite{landau_lifshitz_qm},
\be
R_{E_e l'}(r)
=
\sqrt{\frac{2 m_e}{\pi k_e}}
\frac{2^{l'}}{r}
\frac{(k_e r)^{l'+1}}{(2l'+1)!}
\Gamma(l'+1+i\eta)\, e^{-\pi \eta/2}\, e^{i k_e r}
\,{}_1F_1(l'+1+i\eta,\,2l'+2,\,-2 i k_e r),
\label{eq:R_continuum}
\ee
where \(k_e=\sqrt{2 m_e E_e}\) is the outgoing electron momentum and
\(\eta=-m_e\alpha/k_e\) is the Sommerfeld parameter.
The matrix element factorizes into angular and radial parts,
\be\label{eq:iomir}
\int d^3\mathbf{r}\, \psi^*_{E_e l' m'}(\mathbf{r}) (\mathbf{v}\cdot \mathbf{r}) \psi_{nlm}(\mathbf{r})
= (\mathbf{v}\cdot\hat{z})\, I_\Omega\, I_R,
\ee
with
\be
I_\Omega = \int Y^*_{l' m'}(\theta,\phi) \cos\theta Y_{00}(\theta,\phi)\, d\Omega = \frac{1}{\sqrt{3}}\delta_{m',0},\quad
I_R = \int_0^\infty R^*_{E_e,l'=1}(r) R_{1}(r) r^3 dr.
\label{eq:I_omega_I_R}
\ee
Defining $q_e = m_e v$, the differential ionization probability in hydrogen for the orbital $s$ is obtained utilizing Eqs.~(\ref{eq:amplitude}-\ref{eq:iomir}),
\be
\frac{dP^{(s)}_H}{dE_e} = \frac{q_e^2}{3}\,|I_{R,s}|^2.
\ee
\noindent for $s=sp^2, sp^3$.  Evaluating the radial integral gives,

\be
|I_{R,s}|^2
=
2^8 \left(\frac{a_0}{z_{eff}}\right)^{7} \,
\left(2+\eta k_e \frac{a_0}{z_{eff}}\right)^2
\frac{m_e k^3_e}{\left(1+k_e^2a^2_0/z^2_{eff}\right)^6}
\,
\frac{\eta (1+\eta^2)}{e^{2 \pi \eta} - 1}
\,
e^{4 \eta \,\arctan(k_e a_0/z_{eff})}
\ee
so that the final differential probability reads,

\be\label{eq:iprobh}
\frac{dP_H^{(s)}}{dE_e}
=
\frac{2^8}{3} \left(\frac{a_0}{z_{eff}}\right)^{7} \,
\left(2+\eta k_e \frac{a_0}{z_{eff}}\right)^2
\frac{m_e q^2_e k^3_e}{\left(1+k_e^2a^2_0/z^2_{eff}\right)^6}
\,
\frac{\eta (1+\eta^2)}{e^{2 \pi \eta} - 1}
\,
e^{4 \eta \,\arctan(k_e a_0/z_{eff})}.
\ee
In Fig.~\ref{fig:onization_prob}, the hydrogen differential ionization probability for $q_e = 1\,\mathrm{eV}$ is plotted alongside the carbon ionization probabilities presented in Sec.~\ref{app:iprob_carbon}.

The total ionization probability per LAB molecule for WIMP--hydrogen scattering events is then obtained summing over 5 $sp^2$ orbitals and 21 $sp^3$ orbitals in each LAB molecule, i.e.,

\be
\frac{dP_H}{dE_e} = \left [5\times \frac{dP_H^{(sp^2)}}{dE_e}+21\times  \frac{dP_H^{(sp^3)}}{dE_e} \right ].
\label{eq:prob_tot_H}
\ee

\subsection{Carbon ionization probabilities}
\label{app:iprob_carbon}

In the case of carbon we can distinguish between inner orbitals ($n$ =1), for which the isolated atom approximation can be safely assumed  for the ionization probability (we directly adopt the evaluation in~\cite{Ibe:2017yqa}) and external ones ($n$=2) that are hybridized and affected by molecular bonds.  
In this case, as anticipated, we proceed by assuming orbitals with the Slater--type radial dependence of Eq.~(\ref{eq:slater_orbital}), using the values of $Z_{eff}$ and $E_{ion}$ summarized in Table~\ref{table:C_orbital}. The radial part of the calculation proceeds as in Section~\ref{app:iprob_hydrogen} with the $l^{\prime}$--dependent continuum final state given in~(\ref{eq:R_continuum}) and the result of the radial integration given in Eq.~(\ref{eq:I_omega_I_R}).
As for the angular integration, it requires more care due to the fact that the $sp^3$ and $sp^2$ orbitals are not isotropic and the quantization axis is fixed in the molecular reference frame (and not along the nuclear recoil velocity {\bf v} as in Eq.~(\ref{eq:iomir})). Instead, the angular integration needs to be averaged over the directions of {\bf v} since the orientations of the LAB molecules are randomly distributed. Then, defining the two unit vectors $\hat{n}_{r}(\Omega)$ and $\hat{n}_{v}(\Omega_v)$ that point in the radial direction and in the direction of the nuclear recoil velocity $v$,

\be
I_{\Omega, s}^{l^{\prime}m^{\prime}} = \frac{1}{4\pi} \int d\Omega_v \int d\Omega \, Y_{l', m'}(\Omega) \hat{n}_{v}(\Omega_v) \cdot \hat{n}_{r}(\Omega) \, \Psi_s(\Omega).
\ee

\noindent The hybridized shell $\Psi_s$ are superpositions of $s$ and $p$ states, $\Psi_s=\sum_{l=0,1} \sum_{m=-l}^{l} a_{l,m}Y_{lm}$. Then, in terms of the $a_{l,m}$ coefficients, a straightforward calculation yields,
\be
\left| I_{\Omega, s}^{l'} \right|^2 = \sum_{m^{\prime}}\left| I_{\Omega, s}^{l^{\prime}m^{\prime}} \right|^2 
=\begin{cases} 
\frac{1}{9} \left( \left| a_{1, -1} \right|^2 + \left| a_{1, 0} \right|^2 + \left| a_{1, 1} \right|^2 \right) & \quad l' = 0 \\[1ex]
\frac{1}{3} \left| a_{0, 0} \right|^2 & \quad l' = 1 \\[1ex]
\frac{2}{9} \left( \left| a_{1, -1} \right|^2 + \left| a_{1, 0} \right|^2 + \left| a_{1, 1} \right|^2 \right) & \quad l' = 2,
\end{cases}
\label{eq:I_omega_l_s}
\ee
\noindent with the sum over the  values of $m^{\prime}$ allowed by the dipole selection rule. For the three types of hybridized orbitals the $a_{l,m}$ coefficients are given by~\cite{atkins2011molecular},
\begin{eqnarray}
 \Psi_{sp^2}&=&\sqrt{\frac{1}{3}} Y_{0,0} +\sqrt{\frac{2}{3}} Y_{1,0},\,\left(a_{0,0}=\sqrt{\frac{1}{3}}, a_{1,0}=\sqrt{\frac{2}{3}}\right) \nonumber\\
\Psi_{p_z}&=&Y_{1,0},\qquad\qquad\qquad\left(a_{0,0}=0, a_{1,0}=1\right) \nonumber\\
\Psi_{sp^3}&=&\frac{1}{2} Y_{0,0} +\frac{\sqrt{3}}{2} Y_{1,0},\,
\left(a_{0,0}=\frac{1}{2}, a_{1,0}=\frac{\sqrt{3}}{2}\right), 
\label{eq:hybridized_orbital}
\end{eqnarray}
\noindent with $a_{1,\pm1}$ = 0 when the $z$ axis is taken in the direction of the axis of each orbital. Substituting in Eq.~(\ref{eq:I_omega_l_s}) the $a_{l,m}$ coefficients of Eq.~(\ref{eq:hybridized_orbital}) one obtains 
the $\left| I_{\Omega, s}^{l'} \right|^2$ values provided in Table~\ref{table:I_omega} for each of the $l^{\prime}$'s allowed by the dipole selection rule\footnote{As already pointed out, the ionization of the $p_z$ delocalized orbital may take an additional non--adiabatic contribution~\cite{Blanco:2022pkt} due to its low ionization energy. In particular, such contribution can change the selection rule of the process, implying a non--vanishing parity--conserving contribution for $l^{\prime}$ =1. We neglect such contribution in our analysis}. 

Then, the probability for an electron in the orbital $s$ in a carbon atom is given by,
\be
\frac{dP^{(s)}_C}{d E_e} = q_e^2 \sum_{l^{\prime}}|I_{R,s}^{l^{\prime}}|^2 \left| I_{\Omega, s}^{l'} \right|^2.
\ee

The total ionization probability per LAB molecule for WIMP--carbon scattering events is then obtained summing  over 18 $sp^2$ electrons (3 for each of the 6 carbons in the benzene ring), 40 $sp^3$ electrons (4 in each of the 10 carbons in the Alkyl tail), 4 $p_z(1,1')$ electrons and 2 $p_z(2)$ electrons (in each benzene ring) and 32 $1s$ electrons (2 for each of the 16 carbon atoms), i.e.,

\be
\frac{dP_C}{dE_e} = \left [18\times \frac{dP_C^{(sp^2)}}{dE_e}+40\times  \frac{dP_C^{(sp^3)}}{dE_e}+4\times\frac{dP_C^{(p_z(1,1'))}}{dE_e}+2\times \frac{dP_C^{(p_z(2))}}{dE_e}+32\times \frac{dP_C^{(1s)}}{dE_e}  \right ].
\label{eq:prob_tot_H}
\ee

\begin{table}
\centering
\begin{tabular}{|c|c|c|}
\toprule
bond & $Z_{eff} $ & $E_{ion}$ (eV) \\
\midrule
$sp^2$& 3.6 & 15.5\\
$sp^3$& 3.42 & 14.5\\
$p_z$ ($1'$, $1$) & 3.15 & 9.25\\
$p_z$ ($2$) & 3.15 & 11.5\\
\bottomrule
\end{tabular}
\caption{Slater screening constants $Z_{eff}$ and ionization probabilities for molecular orbitals in carbon targets~\cite{potts1972photoelectron, karlsson1973high,Blanco:2019lrf}.}
\label{table:C_orbital}
\end{table}

\begin{table}
\centering
\begin{tabular}{|cc|c|c|c|}
\toprule
 & $l^{\prime}=0$ & $l^{\prime}=1$ & $l^{\prime}=2$\\
\midrule
$sp^2$& $\frac{2}{27}$ & $\frac{1}{9}$ & $\frac{4}{27}$   \\
$p_z$& $\frac{1}{9}$ & $0$ & $\frac{2}{9}$  \\
$sp^3$& $\frac{1}{12}$ & $\frac{1}{12}$ & $\frac{1}{6}$\\  
\bottomrule
\end{tabular}
\caption{The angular integrals $\left| I_{\Omega, s}^{l'} \right|^2$ defined in Eq.~(\ref{eq:I_omega_l_s}) allowed by the dipole selection rule and averaged over the orientation of the LAB molecule.}
\label{table:I_omega}
\end{table}

\section{Quenching}
\label{app:quenching}

Quenching factor measurements in LAB exist only for recoil energies in the MeV range~\cite{quenching_LAB1, quenching_LAB2}. However, measurements at lower energies are available for liquid scintillators similar to it. In particular, pseudocumene (C$_9$ H$_{12}$) belongs to the same class of alkylated aromatic hydrocarbons as LAB and also a very similar density, as well as a  very similar ratio between the number of carbon and hydrogen targets. In Ref.~\cite{quenching_hong} its scintillation efficiencies for hydrogen (proton) and carbon recoils were measured down to ~30 keV and 45 keV, respectively. Such low--energy measurement were then used in Ref.~\cite{quenching_tetryak} to extrapolate the quenching factor of C$_9$ H$_{12}$ to lower energies using numerically calculated stopping powers off electrons for C and H ions to fit Birks formula~\cite{quenching_birks}. In both cases the quenching factor extrapolated in this way rises at low--energy, although this is a physical effect only in the case of proton recoils, for which scattering off ions can be neglected and the stopping power off electrons spikes at low velocity (Bragg peak). As a consequence, in Fig. 3d of~\cite{quenching_tetryak} the low--energy quenching factor for proton recoils stays around 10\% down to very low energies. In our analysis for H recoils in LAB we adopt the representative value $Q$ =0.1. On the other hand, in the same figure the extrapolation at low--energy of the quenching factor for C recoils is around 5\%, but must be multiplied times the Lindhard partition  function $f_n$~\cite{Lindhard1963} to take into account the stopping power of other C atoms,
\begin{eqnarray}
f_n &=& \frac{k g(\varepsilon)}{1 + k g(\varepsilon)} \\
\varepsilon = 11.5 \, E_R(\text{keV}) \, Z^{-7/3}, &\quad& k = 0.133 \, Z^{2/3} A^{1/2} \\
g(\varepsilon) &=& 3 \, \varepsilon^{0.15} + 0.7 \, \varepsilon^{0.6} + \varepsilon
\end{eqnarray}
\noindent with $E_R$ the recoil energy, and $A$ =12, $Z$ =6.  Since for $E_{ee}<$ 0 keV one has $f_n\ll$ 1 the quenching factor for carbon recoils turns out to be strongly suppressed at low energy. For this reason in our analysis we assume for LAB $Q$ = 0 for carbon. 

\providecommand{\href}[2]{#2}\begingroup\raggedright\endgroup

\end{document}